\documentclass[journal]{elsarticle}

\usepackage{amsmath,amssymb,amsfonts,amstext,amsthm, amsrefs}
\usepackage{comment}
\usepackage[lined,boxruled]{algorithm2e}
\usepackage{ifpdf}
\usepackage{graphicx,amssymb,lineno}
\usepackage{epsfig}
\usepackage[emtex]{color}
\usepackage {eepic}
\usepackage{latexsym}
\usepackage{dsfont}
\usepackage{euscript}
\usepackage{mathrsfs}
\usepackage{graphics}
\usepackage {epic}
\usepackage {eepic}
\usepackage{latexsym}
\usepackage{fancybox}
\usepackage[active]{srcltx}
\usepackage{ulem}

\newtheorem{thm}{Theorem}
\newtheorem{clm}[thm]{Claim}
\newtheorem{defn}[thm]{Definition}
\newtheorem{lem}[thm]{Lemma}
\newtheorem{cor}[thm]{Corollary}
\newtheorem{prop}[thm]{Proposition}

\newcommand{\NP}{{\mathsf{NP}}}

\newcommand{\APX}{{\mathsf{APX}}}
\newcommand{\DTIME}{{\mathsf{Dtime}}}

\begin{document}
\begin{frontmatter}
\title{On the Complexity of Making a Distinguished Vertex Minimum or Maximum Degree by Vertex Deletion}
\author[iitm]{Sounaka Mishra \corref{cor1}} \ead{sounak@iitm.ac.in}
\author[iitmee]{Ashwin Pananjady}  \ead{ee10b025@ee.iitm.ac.in}
\author[iitm]{Safina Devi N}  \ead{safina@smail.iitm.ac.in}
\cortext[cor1]{Corresponding author}
\address[iitm]{Department of Mathematics, Indian Institute of Technology Madras, India, 600036}
\address[iitmee]{Department of Electrical Engineering, Indian Institute of Technology Madras, India, 600036}
\begin{abstract}
In this paper, we investigate the approximability of two node deletion problems. Given a vertex weighted graph $G=(V,
E)$ and a specified, or ``distinguished" vertex $p \in V$, \texttt{MDD(min)} is the problem of finding a minimum weight vertex set $S
\subseteq V\setminus \{p\}$ such that $p$ becomes the minimum degree vertex in $G[V \setminus S]$; and
\texttt{MDD(max)} is the problem of finding a minimum weight vertex set $S \subseteq V\setminus \{p\}$ such that
$p$ becomes the maximum degree vertex in $G[V \setminus S]$. 
These are known $\NP$-complete problems and have been studied from the parameterized complexity point of view in
\cite{betzler}. Here, we prove that for any $\epsilon > 0$, both the problems cannot be approximated within a
factor $(1 - \epsilon)\log n$, unless $\NP \subseteq \DTIME(n^{\log\log n})$. We also show that 
for any $\epsilon > 0$, \texttt{MDD(min)} cannot be approximated within a factor $(1 -
\epsilon)\log n$ on bipartite graphs, unless $\NP \subseteq \DTIME(n^{\log\log n})$, and that for any $\epsilon > 0$,
 \texttt{MDD(max)} cannot be approximated within a factor $(1/2 - \epsilon)\log n$ on bipartite graphs,  unless
$\NP \subseteq \DTIME(n^{\log\log n})$. We give an $O(\log n)$ factor approximation 
algorithm for \texttt{MDD(max)} on general graphs, provided the degree of $p$ is $O(\log n)$.
We then show
that if the degree of $p$ is $n-O(\log n)$,  a similar result holds for  \texttt{MDD(min)}. We prove
that \texttt{MDD(max)} is $\APX$-complete on 3-regular unweighted graphs and provide an approximation algorithm
with ratio $1.583$ when $G$ is a
3-regular unweighted graph. In addition, we show that
\texttt{MDD(min)} can be solved in polynomial time when $G$ is a regular graph of constant degree.
\end{abstract}
\end{frontmatter}

\section{Introduction}
The problems of making a distinguished vertex minimum or maximum degree by vertex deletion in undirected
graphs are very natural, albeit unexplored problems in graph theory, and see a wide array of applications. 
We formally state these two problems as follows.
\begin{itemize}
\item \texttt{MDD(min)}: Given a graph $G=(V, E)$ with a distinguished vertex $p\in V$, find a vertex set $S
\subseteq V\backslash\{p\}$ of minimum size such that the vertex $p$ is the unique vertex of minimum degree in $G[V
\setminus S]$.
\item
\texttt{MDD(max)} : Given a graph $G=(V, E)$ with a distinguished vertex $p$, find a vertex set $S
\subseteq V\backslash\{p\}$ of minimum size such that the vertex $p\in V$ is the unique vertex of maximum degree in $G[V
\setminus S]$.
\end{itemize}
Variants of these problems include the \emph{weighted case}, in which we are interested in finding a vertex set $S$
of minimum weight instead of minimum cardinality, when each vertex in $G$ has a weight associated with it.

These problems have been previously studied in \cite{betzler, betzler1} with reference to directed graphs and
electoral networks.
The most natural motivation lies in competitive social networks, which are undirected, and in which the degree of a
node is
widely seen as a measure of its popularity, influence or importance. An agent may wish to decrease the influence 
of a competing agent (minimize the degree of a distinguished vertex) or increase his own influence (maximizing 
degree of a distinguished vertex) at minimum cost, by shielding the minimum number of other agents from the
network.
 
Another application lies in terrorist networks studied extensively in \cite{terroristnet1, tnet2}, in which the
connectivity of a particular node in the network may be decreased by targeting the minimum number of other nodes.
The \texttt{MDD(min)} problem finds a direct application in this scenario, as well as in similar scenarios
involving cartel networks.
 
The third major application could lie in biology - in protein networks. There have been a multitude of papers
published \cite{protein1,protein2,protein3} which try to correlate the parameter of a particular node in the
network - such as degree, centrality, etc. - with the
importance of the corresponding protein. While degree is seen as a reasonably good indicator of connectivity and
influence,
it may be interesting to look at how many other proteins would have to disappear from the network in order to make
a particular protein
influential. This is a direct application of \texttt{MDD(max)}, and the minimum number of other proteins which
need to be deleted 
could provide a measure of essentiality of the protein corresponding to the distinguished vertex.
The research in this area has been mainly empirical so far, and this could provide another metric  to judge
the importance of a particular protein given its interaction network.

Both \texttt{MDD(min)} and \texttt{MDD(max)} are known to be $\NP$-complete \cite{betzler}. Previous work on these
two problems 
involved approaches using parameterized complexity \cite{betzler}, but a classical complexity approach has not yet
been taken as per
our knowledge. In this paper, we take a classical complexity theory approach towards the problems and make the
following contributions:
\begin{itemize}
\item We show that \texttt{MDD(min)} on a graph $G$ is equivalent to \texttt{MDD(max)} on the graph $G^c$. 
\item We prove that both \texttt{MDD(min)} and \texttt{MDD(max)} are hard to approximate within a factor smaller
than $\log n$, where $n$ represents the number of vertices in the input graph.
\item On bipartite graphs, we prove that \texttt{MDD(min)} and \texttt{MDD(max)} are hard to approximate
within a factor smaller than $O(\log n)$.
\item We propose a $O(\log n)$ factor approximation algorithm for \texttt{MDD(max)} when the input graph $G$
satisfies a certain property. As a consequence of this, we show that if $d(p) = O(\log n)$ in the input
graph $G$, \texttt{MDD(max)} is approximable within a factor of $O(\log n)$.
\item We show that \texttt{MDD(min)} is solvable in polynomial time on $k$-regular graphs, as long as $k=O(1)$.
\item For 3-regular unweighted graphs, we propose an approximation algorithm for \texttt{MDD(max)} with
approximation
ratio $1.583$. On 3-regular bipartite graphs, we prove that \texttt{MDD(max)} is $\APX$-complete.
 
\end{itemize}
\section{Preliminaries}
All the discussion in this paper concerns undirected graphs. The word \emph{graph} is used to mean undirected graph without any ambiguity.
\subsection{Notation} \label{Sec_notation}
In a graph $G=(V,E)$, the sets $N_G(v) = \{u \in V(G): (u, v) \in E\}$ and $N_G[v] = N_G(v)\cup \{v\}$ denote the
\emph{neighbourhood} and the \emph{closed neighbourhood} of a vertex $v$ in $G$, respectively. The
\emph{degree} of a vertex $v$ in $G$ is $|N_G(v)|$ (or the number of neighbours of $v$ in graph $G$, and is denoted by $d_G(v)$. Note that even if $v\not\in V(H)$, $d_H(v)$ could be non-zero, if $v\in V(G)$ and $H$ is a subgraph of $G$. We shall use $N(v)$, $N[v]$ and
$d(v)$ instead of $N_G(v)$, $N_G[v]$ and $d_G(v)$, respectively, when there is no ambiguity regarding the graph under consideration. In a similar vein, for a set of vertices $S$, we define $N(S) = \cup_{v\in S} N(v)$ and $N[S] = \cup_{v\in S} N[v]$.

A graph $G=(V, E)$ is called \emph{$k$-regular} if  $d_G(v)=k$, $\forall$ $v\in V$. For $S \subseteq V$, $G[S]$ denotes the
subgraph induced by $S$ on $G$. The \emph{complement of a graph}  $G=(V, E)$ is the graph $G^c=(V, E^c)$,  where
$(u,v)\in E^c$ if and only if $(u, v)\notin E$, $\forall$ $u, v\in V, u\neq v$. Unless
otherwise mentioned, $n$ denotes the number of vertices in the input graph.

In a graph $G=(V, E)$, $S \subseteq V$ is called a \emph{dominating set} in $G$ if $N[S] = V$.
Given a graph $G=(V, E)$, an instance of \texttt{MDD(max)}, we say that $S \subseteq V\setminus \{p\}$
is a \emph{solution to \texttt{MDD(max)}} for $G$, if the vertex $p$ is the maximum degree vertex in $G[V\setminus
S]$. $S$ is called a \emph{minimal solution to \texttt{MDD(max)}} for $G$ if, for each $u \in S$, $S \setminus
\{u\}$ is not a solution to \texttt{MDD(max)} for $G$. A \emph{minimum solution} to \texttt{MDD(max)} on 
graph $G$ is a solution $S$ to \texttt{MDD(max)} of minimum weight/cardinality. Similarly, a solution (and 
minimal solution, minimum solution) to \texttt{MDD(min)} for $G$ is defined.

\subsection{Known Results}
We now state the definitions of a few known $\NP$-complete optimization problems such as the \emph{minimum dominating set} problem,
\emph{f-dependent set} problem and \emph{minimum set cover} problem, and state approximability and inapproximability bounds for
them.

\begin{defn}[\texttt{MinDom}]
Given a graph $G=(V, E)$, the minimum dominating set problem \texttt{MinDom} is to find a dominating set 
$S$ of minimum cardinality.
\end{defn}

Given a universe $\mathcal{U}=\{x_1,x_2,\ldots, x_r\}$ and a collection of subsets $\mathcal{F}=\{F_1,F_2,\ldots,
F_t\}$ where $F_i \subseteq \mathcal{U}$, a set $T \subseteq \mathcal{F}$ is called a \emph{set cover} for $\mathcal{U}$ if 
$\cup_{F\in T} F=\mathcal{U}$. Size of a set cover $T$ is defined as the number of sets in it.

\begin{defn}[\texttt{MinSetCover}]
Given an instance $(\mathcal{U}, \mathcal{F})$, the minimum set cover problem \texttt{MinSetCover} is to find a
set cover $T$ of minimum size.  \label{minsetcoverdef}
\end{defn}

Both \texttt{MinDom} and \texttt{MinSetCover} are known to be equivalent with respect to approximation 
preserving reductions \cite{kannThesis} and both cannot be approximated within a factor better than $\log n$.

\begin{prop} {\rm \cite{feige}} \label{thmfeige}
For any $\epsilon > 0$, \texttt{MinDom} and \texttt{MinSetCover} cannot be approximated within a factor $(1 -
\epsilon)\log n$, unless $\NP \subseteq \DTIME(n^{\log \log n})$. Note that for \texttt{MinSetCover}, $n=|\mathcal{U}|+|\mathcal{S}|=r+t$.
\end{prop}

Another inapproximability result for \texttt{MinDom} is also known, and we will use it in some of our proofs.
\begin{prop} {\rm \cite{papaYa}} \label{mindomapx}
\texttt{MinDom} is $\APX$-complete for cubic (3-regular) as well as bicubic (3-regular bipartite) graphs.
\end{prop}

\begin{defn}[f-dependent set deletion]
Given a vertex weighted graph $G=(V, E)$ and a function $f:V \rightarrow \mathbb{N}$, the $f$-dependent set deletion
problem is to find a set $S \subseteq V$ of minimum weight such that degree of each vertex $v$ in $G[V \setminus S]$
is at most $f(v)$.
\end{defn}

\begin{prop} {\rm \cite{okunbarak}} \label{thmOB}
The $f$-dependent set problem can be approximated within a factor of $2 + \log \alpha$, where $\alpha =
\mbox{max}\{f(v)|v \in V\}$ and $f(v) \geq 3$ for all $v \in V$.
\end{prop}

The $f$-dependent set problem is a generalization of \texttt{MinDom} and has a similar inapproximability result which is as follows.

\begin{prop} {\rm \cite{okunbarak}} \label{thmOBL}
Unless $\NP \subseteq \DTIME(n^{\log \log n})$, for any $\epsilon > 0$, $f$-dependent set problem cannot be
approximated within a factor of $(1 - \epsilon)\log \alpha$, where $\alpha = \mbox{max}\{f(v)|v \in V\}$ and $f(v)
\geq 3$ for all $v \in V$.
\end{prop}

\subsection{Equivalence of \texttt{MDD(min)} and \texttt{MDD(max)}}
We now show a result that we will use repeatedly in this paper.

\begin{thm} \label{max1}
\texttt{MDD(max)} in a graph $G$ is equivalent to \texttt{MDD(min)} in graph $G^c$, and vice versa.
\end{thm}
\noindent
{\bf Proof :} Given an instance $G=(V, E)$ of \texttt{MDD(max)}, we construct the graph $G^c$ as an
instance of \texttt{MDD(min)}. An optimal solution to  \texttt{MDD(max)} for $G$ for \texttt{MDD(max)} 
must be an optimal solution to \texttt{MDD(min)} for $G^c$, since the two operations - deletion and complementation are
commutative as far as our problem is concerned. From this observation, the theorem statement follows. \qed

From Theorem \ref{max1}, it also follows that both \texttt{MDD(min)} and \texttt{MDD(max)} are equivalent with
respect to approximation preserving reductions.

\section{Hardness Results} \label{sec_hardness}
In this section, we show that both \texttt{MDD(min)} and \texttt{MDD(max)} are hard to approximate within a 
factor smaller than $O(\log n)$. We prove these results by establishing approximation preserving reductions from
\texttt{MinDom} and  using the inapproximability result from Proposition \ref{thmfeige}.

\begin{thm} \label{min1}
For any $\epsilon > 0$, \texttt{MDD(min)} cannot be approximated within a factor $(1 - \epsilon)\log n$, unless
$\NP \subseteq \DTIME(n^{\log \log n})$.
\end{thm}
\noindent
{\bf Proof :} 
Given an instance $G=(V, E)$ of \texttt{MinDom}, we construct an instance $H=(V', E')$ of \texttt{MDD(min)} in
polynomial time, as follows. Here, we assume that $n$ is the number of vertices in $G$. First, we construct the
complement $G^c$ of $G$. Then, we create a new vertex $p$ and join it to
all the vertices in $V$ by introducing $n$ edges $(p, v)$ $\forall$ $v \in V$. Next, we add a complete graph
$K_{2n+2}$ over a set $T$ of $2n+2$ new vertices. For each vertex $v \in V$, if the degree of $v$ is $x$ in
$G$, i.e. $d_G(v)=x$, we add $x$ edges from $v$ to any $x$ vertices of $T$, to form graph $H$. Notice that now,
$\forall$ $v\in G$, $d_H(v)=n$, due to the complementation of $G$ in the construction of $H$. It is easy to observe that $H$ has
$3n+3$ vertices as $V' = V \cup \{p\} \cup T$.
\begin{figure}[h]
\begin{center}
    \includegraphics[scale=.5]{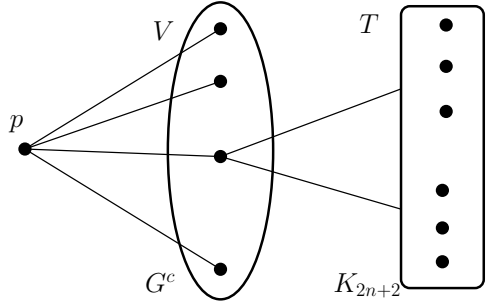}
\end{center}
\caption{Construction of $H$}
\end{figure}

We now claim that:
\begin{clm} \label{clm_1}
$S \subseteq V$ is a dominating set in $G$ if and only if $p$ is the vertex of minimum degree
in $H[V' \setminus S]$ (i.e. $S$ is a solution to \texttt{MDD(min)} for $H$).
\end{clm}
\noindent
{\bf Proof :}
Let $S \subseteq V$ be a dominating set in $G$. Then for all $v \in V \setminus S$, $v$ is adjacent to
some vertex
in $S$. Therefore, the degree of a vertex $v$ in $H[V' \setminus S]$ is at least $n - |S| + 1$. At the same time,
the degree
of $p$ in $H[V' \setminus S]$ is $n-|S|$. The degree of each vertex in $V' \setminus  (V\cup \{p\})$ at least
$2n+1$, by construction.
Since degree of $p$ in $H[V' \setminus S]$ is $n-|S|$, it follows that $p$ is the minimum degree vertex in
$V'\setminus S$, and therefore, $S$ is a solution to \texttt{MDD(min)} on $H$.

Conversely, let $S \subseteq V' \setminus \{p\}$ be a vertex deletion set in $H$ which makes $p$ the vertex of minimum degree in $H[V'\setminus S]$. Since $|T| = 2n+2$ and all
vertices in $T$ have large degree, the optimal
vertex deletion set in $H$ cannot have size larger than $|V|$. Therefore, an optimal vertex deletion set in $H$ is
a subset of $V$. Based on this observation, we shall assume that any vertex deletion set $S$ in $H$ is a subset of
$V$. Since $S$ is a vertex deletion set, $d_{H[V' \setminus S]}(p) = n-|S|$ and $d_{H[V' \setminus S]}(u) \geq
n-|S|+1$, for all $u \in V' \setminus (S\cup\{p\})$. Let $v \in V \setminus S$ be any vertex. Since $d_{H[V'
\setminus S]} \geq n-|S|+1$, there exists at least one vertex $u \in S$ such that $v$ and $u$ are not adjacent in
$H$. This implies that $S$ is a dominating set in $G$.
\qed

From Claim \ref{clm_1}, it follows that the reduction explained is a cost preserving reduction. 
Since $|V'| = 3(n+1)$, which is linear in $n$, and using Proposition \ref{thmfeige}, it can be observed that for
sufficiently large $n$ and for any $\epsilon'>0$, \texttt{MDD(min)} cannot be approximated within a factor of  $(1
- \epsilon')\log |V'|$, unless $\NP \subseteq \DTIME(n^{\log \log n})$. Theorem \ref{min1} is therefore proved. \qed

Using Theorem \ref{max1}, it follows as a corollary that:
\begin{cor}
For any $\epsilon > 0$, \texttt{MDD(max)} cannot be approximated within a factor $(1 - \epsilon)\log n$, unless
$\NP \subseteq \DTIME(n^{\log \log n})$.
\end{cor}
We now prove that a similar hardness result holds for \texttt{MDD(min)} even when the input $G$ is restricted to
bipartite graphs. We do this by establishing a cost preserving reduction from \texttt{MinSetCover} 
to \texttt{MDD(min)} on bipartite graphs, similar to that of Theorem \ref{min1}. 

\begin{figure}[h]
\begin{center}
    \includegraphics[scale=.50]{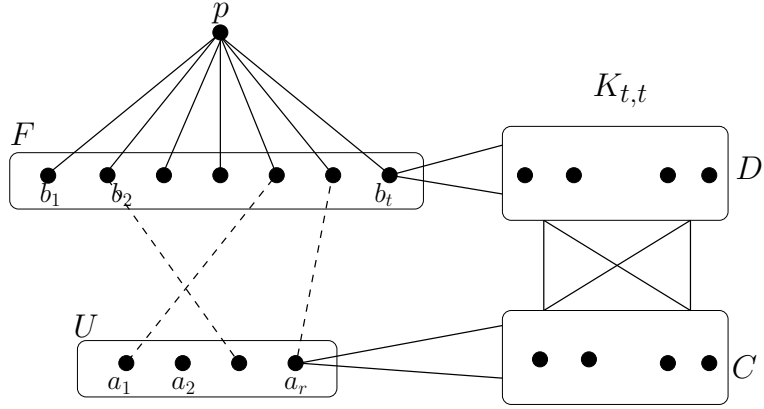}
\end{center}
\caption{\texttt{MDD(min)} for bipartite graphs: Construction of the graph $G$ from an instance  $(\mathcal{U},
\mathcal{F})$ of set cover}
\label{bipartite}
\end{figure}

\begin{thm} \label{min2}
For any $\epsilon > 0$, \texttt{MDD(min)} on bipartite graphs cannot be approximated within a factor $(1 -
\epsilon)\log n$, unless $\NP \subseteq \DTIME(n^{\log \log n})$.
\end{thm}
\noindent
{\bf Proof :} We prove this theorem by establishing a cost preserving reduction from \texttt{MinSetCover} 
to \texttt{MDD(min)}. Let $(\mathcal{U}, \mathcal{F})$ be an instance of \texttt{MinSetCover} with  
$\mathcal{U}=\{x_1, x_2, \ldots, x_r\}$, $\mathcal{F} =\{F_1, F_2, \ldots, F_t\}$ (refer Definition \ref{minsetcoverdef}). Here we assume that 
$|\mathcal{U}| < < |\mathcal{F}|$. We construct  a graph $G=(V, E)$ corresponding 
to $\mathcal{U}$ and $\mathcal{F}$ as follows. We introduce a
vertex for every element in $\mathcal{U} \cup \mathcal{F}$.
Let $U = \{a_1, a_2, \ldots, a_r\}$ be  the set of vertices corresponding to elements in $\mathcal U$, where
vertex $a_i$ corresponds to element $x_i\in \mathcal{U}$ and 
$F=\{b_1, b_2, \ldots, b_t\}$ be the set of vertices  corresponding to the elements in $\mathcal{F}$, where
vertex $b_i$ corresponds to subset $F_i$. The vertex set of $G$, $V=U \cup F \cup C \cup D \cup \{p\}$, where $C$ and $D$ have $t$ vertices each. Therefore $|V|=3t+r+1$. 
The edge set $E$ is defined as follows. We make a complete bipartite graph $K_{t,t}$ on $C\cup D$ with vertex
bipartition as $C$ and $D$. We introduce an edge $(p, a_i)$, for every $a_i \in F$. We add an edge 
$(a_i, b_j) \in E$ if and only if $x_i \notin F_j$. At this stage, if the degree of a vertex $b_i \in F$ 
is strictly less than $t$, then we add sufficient edges from $b_i$ to vertices in $D$ in order to increase
$d_G(b_i)$ to $t$. Note that $d_G(b_j)\geq t$ $\forall$ $b_j\in F$. Similarly, we add edges from each vertex $a_i \in U$ to vertices in $C$
such that $d_G(a_i)\geq t$ $\forall$ $a_i\in U$. Clearly, $G$ is a bipartite graph. For a sketch of $G$ see  Figure \ref{bipartite}.

We now prove the following claim.
\begin{clm} \label{clm_2}
$T=\{F_{i_1}, F_{i_2}, \ldots, F_{i_\ell}\}$ is a set cover for $\mathcal U$ if and only if 
$S= \{b_{i_1}, b_{i_2}, \ldots, b_{i_\ell}\}$ is a solution to \texttt{MDD(min)} for $G$.
\end{clm}
\noindent
{\bf Proof :} Note that $d_{G[V\setminus S]}(p)=t-\ell$  and $d_{G[V\setminus S]}(b_i)\geq t$, $\forall$ $b_i\in
F\setminus S$. If $T$ is a set cover 
for $\mathcal{U}$, then for each $k \in \{1, \ldots, r\}$, $\exists j \in \{1, 2, \ldots, \ell\}$ such
that  $x_k\in F_{i_j}$. This implies that for every $a_k \in U$, $\exists j \in \{1, 2, \ldots , \ell\}$
such
that  $(a_k, b_{i_j}) \notin E$ and therefore, $d_{G[V\setminus S]}(a_k) \geq t-\ell+1$,
$\forall$ $a_k\in U$. Also, $d_{G[V\setminus S]}(v)\geq t$ for every $v\in C\cup D$. Thus, $p$ is the unique 
minimum degree vertex in $G[V\setminus S]$, and so $S$ is a solution to \texttt{MDD(min)} for $G$.

Next, we show that for a given minimal solution $S \subseteq V$ to \texttt{MDD(min)} for $G$, we can
construct a set cover $T$ for $\mathcal{U}$ with $|T| \leq |S|$. 
Let $S\subseteq V$ be a minimal solution to \texttt{MDD(min)} for $G$. Then $S \cap F \ne \emptyset$,
since we need to necessarily reduce the degree of $p$. This implies that $d_{G[V\setminus S]}(p) \leq t-1$. It is intuitive that
$S \cap (C \cup D) = \emptyset$. Suppose $S \cap U \ne \emptyset$. Let $a_k \in S \cap U$ be any arbitrary vertex.
Then, for the corresponding element $x_k \in \mathcal{U}$ there exists a set $F_j \in \mathcal{F}$ such that $x_k
\in F_j$. Based on this property, we construct a new set $S'$ of vertices by replacing each vertex $a_k \in S
\cap U$ by a vertex $b_j$ (where $x_k \in F_j$). Therefore, it follows that $|S'| \leq |S|$ and
$S' \subseteq F$. Now we show that the set $T = \{F_i | b_i \in S'\}$ is a set cover for $(\mathcal{U},
\mathcal{F})$. 
If $a_k\in S\cap U$, then by construction of $S'$, there exists an $F_j\in T$ such that $x_k\in \mathcal{U}$.
For $a_k\in U\setminus S$, we have  $d_{G[V\setminus S]}(a_k) > t- |S\cap F|$. This implies that there 
exists at least one $b_j \in S\cap F$ such that $(a_k, b_j)\notin E$. From the construction of $G$, we have that 
$x_k \in F_j$. Note that $S \cap F \subseteq S'$ and therefore $F_j\in T$. Hence, $T$ is a set cover for 
$\mathcal{U}$.
\qed

The reduction in Claim \ref{clm_2} is cost preserving.
Since $|V| = O(n)$, using Proposition \ref{thmfeige}, it can be proved that for any $\epsilon>0$, \texttt{MDD(min)} on bipartite
graphs cannot be approximated within a factor of  $(1 - \epsilon)\log |V|$, unless $\NP \subseteq 
\DTIME(n^{\log \log n})$. Theorem \ref{min2} is therefore proved. \qed 

\begin{figure}[h]
\begin{center}
    \includegraphics[scale=.50]{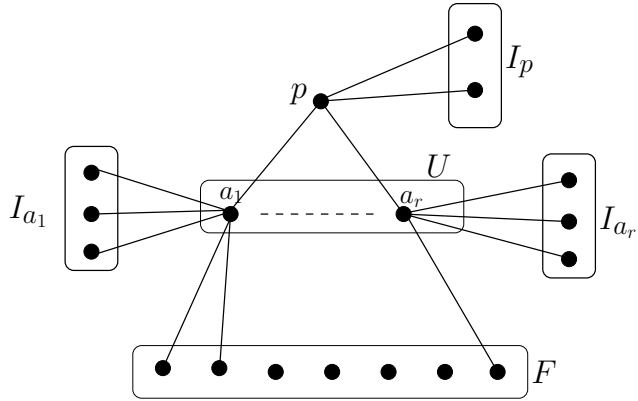}
\end{center}
\caption{\texttt{MDD(max)} for bipartite graphs: Construction of the graph $G$ from an instance $(\mathcal{U}, \mathcal{F})$ of set cover}
\label{bipartitemax}
\end{figure}

Note that the complement of a bipartite graph is not necessarily bipartite, and so Theorem \ref{max1} cannot be used to extend Theorem \ref{min2} to \texttt{MDD(max)} on bipartite graphs. We use a different reduction to show the hardness of \texttt{MDD(max)} on bipartite graphs.
\begin{thm} \label{maximum}
For any $\epsilon > 0$, \texttt{MDD(max)} on bipartite graphs cannot be approximated within a factor
$(\frac{1}{2} - \epsilon)\log n$, unless $\NP \subseteq \DTIME(n^{\log \log n})$.
\end{thm}
\noindent
{\bf Proof :} We prove this theorem by establishing a cost preserving reduction from \texttt{MinSetCover}.
 Let $\mathcal{U} = \{x_1, x_2,\ldots, x_r\}$, $\mathcal{F} =\{F_1, F_2, \ldots, F_t\}$  and 
$|\mathcal{U}| << |\mathcal{F}|$. We
construct a bipartite graph $G$ as follows. First, we construct the natural bipartite graph representation of $(\mathcal{U},\mathcal{F})$.
For this we introduce two sets of vertices as $U = \{a_1, a_2, \ldots, a_r\}$ and $F = \{b_1, b_2, \ldots, b_t\}$,
corresponding to elements in $\mathcal{U}$ and $\mathcal{F}$, respectively. Here, $(a_i, b_j)$ is an edge iff $x_i
\in F_j$. In the next step, we introduce a new vertex $p$ and edges $(p , a_i)$, for $1 \leq i \leq r$. We shall
denote the resulting graph as $G'=(U \cup F \cup \{p\}, E')$. In the final step of the construction of $G$, we
introduce a few degree one vertices to $G'$ so that $d_{G}(v) = t$, for each vertex $v\in U \cup \{p\}$. We do this as follows.
For each $v\in U\cup \{p\}$, we introduce a new set of vertices $I_v$ of size $t-d_{G'}(v)$ to the graph $G'$ and
make $v$ adjacent to all the vertices in $I_v$. Let $I=\cup_{v\in U\cup\{p\}}I_v$.
We call the graph finally obtained as $G=(V,E)$ where $V=U\cup F\cup I\cup \{p\}$ and $E$ is the 
set of edges as defined above. We have that $|V| \leq (|\mathcal{U}|+2)(|\mathcal{F}|+1) \leq n^2$, where $n = r
+ t$. We also observe that $G$ is a bipartite graph, $d(v)=t$ for all $v\in U\cup\{p\}$ and $d(v)<t$ for every
other vertex. For a sketch of this construction, refer to Figure \ref{bipartitemax}.

We now make the following claim.
\begin{clm} \label{clm_3}
$T = \{F_{i_1}, F_{i_2}, \ldots, F_{i_l}\}$ is a set cover for $\mathcal{U}$ if and only if
$S=\{b_{i_1}, b_{i_2}, \ldots, b_{i_l}\}$ is a solution to \texttt{MDD(max)} on $G$.
\end{clm}

\noindent
{\bf Proof:} 
For each $x_i\in \mathcal{U}$ there is an $F_{i_k}\in T$ such that $x_i \in F_{i_k}$ 
and the corresponding vertex $b_{i_k} \in S$. This implies that 
$d_{G[V\setminus S]}(a_i)\leq t-1$, $\forall$ $a_i \in U$ and $d(p)=t$ (since none of the neighbours of $p$ is
deleted). Hence $p$ is the vertex of maximum degree in $G[V\setminus S]$ ($S$ is a solution to \texttt{MDD(max)} on $G$. 

For the converse, let $S$ be a minimal solution to \texttt{MDD(max)} on $G$.
Without loss of generality, we can assume that $S\subseteq F$. Suppose $S\nsubseteq F$ and $v \in S\setminus F$ be
any vertex. If $v=a_i\in U$ for some $i$, then we choose a set $F_j \in \mathcal{F}$ with $x_i \in F_j$ and replace
$v$ by $b_j$ in
$S$. If $v \in I_p$ then we simply remove $v$ from $S$, if $v \in I_{a_k}$, for some $a_k \in U$, then we replace
$v$ by $b_j$ in $S$, where $F_j$ contains the element $x_k$. It is important to observe that this process of
normalizing $S$ does not increase its size. The sub-collection $T =\{F_i : b_i \in S\}$ corresponding to the
vertices of $S$ gives a set cover for $\mathcal{U}$.
\qed

From Claim \ref{clm_3}, it is easy to observe that for any solution $S$ to \texttt{MDD(max)} for $G$, we can find
(in polynomial time) a set cover $T$ for $\mathcal{U}$ with $|T| \leq |S|$. 
Also, if $S_{opt}$ and $T_{opt}$ are
any optimal solutions for \texttt{MDD(max)} and \texttt{MinSetCover}, respectively, then $|S_{opt}| = |T_{opt}|$.
Suppose, for some $\epsilon > 0$, there exists a polynomial time algorithm approximating \texttt{MDD(max)} within
a factor of $(\frac{1}{2} - \epsilon) \log N$, on bipartite graphs with $N$ vertices. Let $S$ be such an
approximate solution to \texttt{MDD(max)} for the bipartite graph $G$ as constructed from an instance
$(\mathcal{U}, \mathcal{F})$ of \texttt{MinSetCover}. Therefore, $|S| \leq |S_{opt}|(\frac{1}{2} - \epsilon) \log
|V|.$ By the above discussion we have, $|T| \leq |S| \leq |T_{opt}|(1 - 2\epsilon) \log n = |T_{opt}|(1 - \epsilon')
\log n$, for some $\epsilon' > 0$. This contradicts Proposition \ref{thmfeige}. Therefore, Theorem \ref{maximum} is proved. \qed


We now consider the complexity of \texttt{MDD(min)} on regular graphs. 
\begin{thm} \label{min-k-reg}
\texttt{MDD(min)} on $k$-regular graphs is solvable in polynomial time as long as $k=O(1)$.
\end{thm}
\noindent
{\bf Proof : }Let $G=(V, E)$ be a $k$-regular graph. We claim that the size of an optimal solution to
\texttt{MDD(min)} for the instance $G$ is at most $2k-1$. We prove this by exhibiting a feasible solution to \texttt{MDD(min)} on $G$ of
size at most $2k-1$.
Let $A=\{v\in V\setminus N[p]:N(v)=N(p)\}$ and let $S=N(p)\cup  A$. We have that $|N(p)|=k$ and
$|A|\leq (k-1)$ since $d(v)=k$ for all $v\in V$. Therefore, $|S|\leq k+(k-1)=2k-1$. Now, consider some vertex
$v\in V\setminus (S\cup\{p\})$. Then $N(v)\neq N(p)$ since $v\not\in A$. Also note that $N(v)\cap A=\phi$. This implies
that $v$ has at least one neighbour in $V\setminus (S\cup \{p\})$. Therefore, $d_G[V\setminus S](v)\geq 1$
$\forall$ $v\in V\setminus (S\cup\{p\})$. Also, $d_{G[V\setminus S]}(p)=0$. Since $d_{G[V\setminus S]}(v)\geq 1$ for
every $v\in V\setminus S$, $p$ is the minimum degree vertex in $G[V\setminus S]$ and hence $S$ is a feasible solution 
to \texttt{MDD(min)} on $G$. Therefore, the size of an optimal solution  to $G$ is at most $2k-1$. 
Let $\mathcal{A}$ be the collection of all subsets of $V\setminus \{p\}$ of size at most $2k-1$.
Then any optimal solution belongs to this collection $\mathcal{A}$. 
We have that $|\mathcal{A}|=\sum_{i=1}^{2k-1}{n \choose i}$ and if this is polynomial in $n$, then an optimal 
solution to $G$ can be found in polynomial time by explicit enumeration of all possibilities. $\sum_{i=1}^{2k-1}{n
\choose i}\approx 2^{nH_2(\frac{2k-1}{n})}$ which is a polynomial in $n$ as long as  $k=O(1)$\footnote{$H_2(x)= -(x\log_2
x+(1-x)\log_2 (1-x))$, $\forall$ $x\in [0,1]$.}.
Therefore, in this case, the optimal solution can be found in polynomial time.
\qed

From Theorem \ref{max1} and Theorem \ref{min-k-reg}, it can be observed that \texttt{MDD(max)} is polynomial time
solvable on $k$-regular graphs provided $k = n - O(1)$. However, we shall prove that \texttt{MDD(max)} on
$k$-regular graphs is $\APX$-complete when $k=O(1)$.

\begin{figure}[h]
\begin{center}
    \includegraphics[scale=.50]{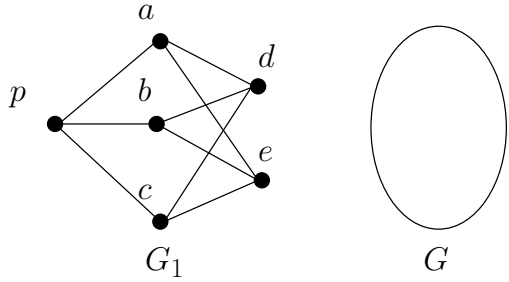}
\end{center}
\caption{Construction of $G'=G\cup G_1$}
\label{cubic1}
\end{figure}

\begin{thm} \label{max2}
\texttt{MDD(max)} is $\APX$-complete on cubic  graphs.
\end{thm}
\noindent
{\bf Proof : }We exhibit a simple {\it L-reduction} \cite{papaYa} from \texttt{MinDom} on cubic  graphs to 
\texttt{MDD(max)} on cubic graphs. Consider a cubic  graph $G=(V, E)$ and an instance of \texttt{MinDom}.
Let $G_1$ be the graph on 6 vertices $\{p, a, b, c, d, e\}$, as given in Figure \ref{cubic1}, and let 
$V'=V\cup \{p, a, b, c, d, e\}$. We construct an instance $(G'=(V', E'), p)$ of \texttt{MDD(max)}, where 
$G'=G_1\cup G$. Clearly, $G'$ is a cubic graph. It is easy to see that the optimal solution to \texttt{MDD(max)} 
for the instance $(G_1, p)$ is the set $\{d, e\}$. This implies that any minimal solution to \texttt{MDD(max)}
for $G'$ contains both $d$ and $e$, and none of $\{a, b, c\}$. Now, to find a solution  for 
$G'$ we only need to bound the degree of every vertex in $G$ by 2. 

If $S$ is a dominating set for $G$, then $d_{G[V\setminus S]}(v)\leq 2$ for every $v\in V\setminus S$. 
Therefore, $S'=S\cup\{d, e\}$ is a solution to \texttt{MDD(max)} for $G'$ with $|S'|=|S|+2$.
Conversely, let $S'$ be a minimal solution  to \texttt{MDD(max)} for $G'$. Then $S'\cap \{a, b, c\}=\emptyset$ 
and $\{d, e\}\subseteq S'$. This implies that $d_{G'[V'\setminus S']}(p)=3$ and $d_{G'[V'\setminus S']}(v)\leq 2$
 for every $v\in V'\setminus S'$ and hence also $d_{G[V\setminus S']}(v)\leq 2$ for every 
 $v\in V\setminus S'$. Thus, $S'\setminus \{d,e\}$ is a  dominating set for $G$ and $|S'|=|S|+2$. 
 
 If $S_{opt}$ is a minimum dominating set for $G$, then $S_{opt}\cup\{d,e\}$ is a minimum solution 
 to \texttt{MDD(max)} for $G'$. Conversely, if $S'_{opt}$ is a minimum solution to \texttt{MDD(max)} for $G'$, 
 then $S'_{opt}\setminus \{d,e\}$ is a minimum dominating set for $G$. Choosing $\alpha=2$, we 
 have that $S'_{opt}\leq \alpha S_{opt}$. Let  $S'$ be a minimal solution to \texttt{MDD(max)} for $G'$
 and let $S$ be the corresponding solution to \texttt{MinDom} for $G$. Then for $\beta=1$, we have that
 $|S|-|S_{opt}|\leq \beta (|S'|-|S'_{opt}|)$. This gives an {\it L-reduction} from  \texttt{MinDom} 
on cubic  graphs to \texttt{MDD(max)} on cubic graphs. From Proposition \ref{mindomapx}, we see that \texttt{MDD(max)} for cubic graphs is $\APX$-hard. In the next
section, we provide a constant factor approximation algorithm to \texttt{MDD(max)} on cubic graphs,
thereby showing that it is $\APX$-complete. \qed

We also arrive at the following Theorem for bicubic (3-regular bipartite) graphs, by a construction similar to that of Theorem \ref{max2}. Note that graph $G_1$ in that construction is bipartite, and so for a bipartite graph $G$, $G'=G_1\cup G$ would be bipartite.
\begin{thm}
\texttt{MDD(max)} is $\APX$-complete for bicubic graphs.
\end{thm}
\noindent
{\bf Proof : }The reduction is similar to that of Theorem \ref{max2}. The constant approximation ratio comes from
an algorithm we present in the next section for cubic graphs. \qed


\section{Approximation Algorithms}
In this section, we show that the vertex weighted version of \texttt{MDD(max)} is approximable within a factor of 
$O(\log n)$, on graphs for which the neighbourhood of vertex $p$ satisfies a particular property. Using Theorem \ref{max1}, we will extend these algorithms to \texttt{MDD(min)}. Here we shall assume that $d(p) = t$ in the input
instance $G=(V, E)$ of \texttt{MDD(max)}. We define $Y=\{v | v \in V \mbox{~and~} d(x) \geq t\}$ and $D = N[Y]$. We will first provide approximation algorithms for special cases of the problem in Lemmas \ref{claim1} and \ref{claim2}, when $Y \cap N[p] = \emptyset$, and then move on to a generalization that captures the aforementioned property even when $Y \cap N[p] \neq \emptyset$.

\begin{lem} \label{claim1}
If the input instance $G$ for \texttt{MDD(max)} satisfies the condition $D \cap N[p] = \emptyset$ then
it can be approximated within a factor of $2 + \log t.$
\end{lem}
\noindent
{\bf Proof : } Consider the $f$-dependent set problem with input as $G[V\setminus N[p]]$ and $f(v)=t-d_{N(p)}(v)
-1$, for all $v \in V \setminus N[p]$. Let $S$ be an approximate solution to the $f$-dependent set problem, for
this instance,
generated by Okun-Barak Algorithm \cite{okunbarak}. We shall show that $S$ is a $(2 + \log t)$-factor approximate
solution of
\texttt{MDD(max)}, for the instance $G$. From the definition of $f$ on $V \setminus N[p]$, it follows that vertex
$p$ is the vertex of maximum degree in $G[V \setminus S]$. Therefore, $S$ is a vertex deletion for
\texttt{MDD(max)} for the instance $G$. Next, we prove that any minimum solution $S_o$ to
\texttt{MDD(max)} for the instance $G$, $S_o \cap N(p)= \emptyset$. Suppose, $A = S_o \cap N(p) \neq \emptyset$.
Let $S'_o =
S_o \setminus A$. Then $S'_o$ is also a vertex deletion set. In the process of deleting the vertices of
$A$ from $S_o$, we increase the degree of vertex $p$ by $|A|$ and vertices in $N(A) \cap [V \setminus (S'_o)]$ by
at most $|A|$. Since degree of each vertex in $A$ is at most $t-1$, it follows that $p$ has maximum degree in
$G[V \setminus S'_o]$. \qed

\begin{lem} \label{claim2}
If the input instance $G$ for \texttt{MDD(max)} satisfies the conditions $Y \cap N[p] = \emptyset$ and $D \cap N[p]
\neq \emptyset$ then it can be approximated within a factor of $2 + \log t.$
\end{lem}
\noindent
{\bf Proof : } Similar to the proof of Lemma \ref{claim1}. Note here that $\forall$ $v:v\in D\cap N[p]$, $v$ will never be part of the solution to \texttt{MDD(max)}. \qed

We are now interested in a more general (but not the most general) case, when $Y \cap N[p] \neq \emptyset$.
Let $G=(V\cup \{p\}, E)$ be an instance of \texttt{MDD(max)} with $Y \cap N(p) \neq \emptyset$. For such an
instance we
construct a set $L \subseteq N(p)$ as given below.

\noindent
\begin{minipage}{1.\linewidth}
\begin{algorithm}[H]
\KwIn{A graph $G=(V, E)$ and $p \in V$ with $Y \cap N(p) \neq \emptyset$\;}
\KwOut{$L \subseteq N(p)$\;}
$L = \emptyset$\;
\While{$\exists$ a vertex $u \in (N(p)\setminus L)$ with $|N(u) \setminus L| \geq |N(p) \setminus L|$}{
 $L = L\cup\{u\}$\;
}
return($L$)\;
\caption{\textbf{Construction of the set $L$}}
\label{algL1}
\end{algorithm}
\end{minipage}

\begin{thm} \label{claim3}
Let $G$ be an instance of \texttt{MDD(max)} with $|L| = O(\log n)$. Then \texttt{MDD(max)} can be approximated
within a factor of $O(\log n)$.
\end{thm}

\noindent
{\bf Proof :} From the definition of $L$ it follows that for every vertex $v \in N(p) \setminus L$, $d(v) < d(p)
\setminus |L|$.  Let $S$ be any solution to \texttt{MDD(max)} for $G$. Then $d_{G[V\setminus S]}(p) > 
d_{G[V\setminus S]}(u)$, for all $u \in V\setminus S$. 

Next we show that any minimal vertex deletion set $S$ in $G$ does not contain any vertex from $N(p)\setminus L$.
Suppose $S \cap (N(p)\setminus L) = A \neq \emptyset$. Let $|A| = \alpha$. Now consider
the set $S' = S \setminus A$. We show that $S'$ is a vertex deletion set. Since, $d_{G[V \setminus S]}(p) > 
d_{G[V \setminus S]}(u)$, for all $u \in V\setminus S$, $d_{G[V \setminus S']}(p) = d_{G[V \setminus S]}(p) + 
\alpha > d_{G[V \setminus S']}(u) = d_{G[V \setminus S]}(u) + \alpha$, for all $u \in V \setminus S$. As 
$d(v) < d(p)-|L|$, for all $v \in A$, we have $d_{G[V \setminus S']}(p) > d_{G[V \setminus S']}(v)$,
for all $v \in A$. 

\noindent
\begin{minipage}{1.\linewidth}
\begin{algorithm}[H]
\KwIn{A graph $G=(V, E)$, $p \in V$ with $Y \cap N(p) \neq \emptyset$ and $|L| = O(\log n), w:V\rightarrow
\mathbb{Z}^+$\;}
\KwOut{A vertex deletion set $S$ for \texttt{MDD(max)} on $G$\;}
$S = \emptyset$\;
$wt = \infty $\;
\For{each subset $K$ of $L$}{
Compute a $f$-dependent set $S'$ using Okun-Barak's algorithm \cite{okunbarak} with input as $G[V\setminus K]$, $w'(v) = \left\{
\begin{array}{ll}
\infty & \mbox{for~} v \in N(p) \setminus K \\
w(v) & \mbox{for~} v \in V \setminus N(p),
\end{array} \right.$
and $f(v) = d(p) - |K| -1$ for $v \in V \setminus K$ \;
$S' = S' \cup K$\;
\If{$w(S') < wt$}{$S = S'$ and $wt = w(S')$\;}
}
return($S$)\;
\caption{\textbf{Computation of  $O(\log n)$ factor solution $S$ for \texttt{MDD(max)}}}
\label{algL2}
\end{algorithm}
\end{minipage}

From the above arguments it follows that any optimal vertex deletion set $S_o$ in $G$ does not contain any vertex
from $N(p) \setminus L$.

Algorithm \ref{algL2} that computes a $O(\log n)$-factor solution to \texttt{MDD(max)}
for the input instance $G$ with $|L| = O(\log n)$. 

Since $|L|=O(\log n)$, Algorithm \ref{algL2} runs in polynomial time.  Let $K_o = S_o \cap L$. Let $S_{K_o}$ 
be the $f$-dependent set  computed in Algorithm \ref{algL2} for the set $K_o$. It is not hard to observe that 
$w(S_o \setminus K_o) = w(S_{o,f,K_o})$, where $S_{o,f,K_o}$ is an optimal $f$ dependent set for the instance
considered in the algorithm associated with set $K_o$. Since the algorithm is choosing the least weight 
vertex deletion set, we have
\small
\begin{align}
\frac{w(S)}{w(S_o)} \leq  \frac{w(K_o) + w(S_{K_o})}{w(S_o)} 
			= \frac{w(K_o) + w(S_{K_o})}{w(K_o) + w(S_o \setminus K_o)} 
			\leq \frac{w(S_{K_o})}{w(S_o \setminus K_o)} 
			&= \frac{w(S_{K_o})}{w(S_{o,f,K_o})} \nonumber \\
			&\leq O(\log n). \nonumber 
\end{align}
\normalsize
\qed

\begin{thm} \label{hardness_L}
For any $\epsilon > 0$, \texttt{MDD(max)} cannot be approximated within a factor $(1 - \epsilon)\log n$, unless
$\NP \subseteq \DTIME(n^{\log \log n})$, even on graphs with $L = \emptyset$.
\end{thm}

\noindent
{\bf Proof :} Follows from Theorem \ref{min1} and Theorem \ref{max1}. Note that in the reduction in the proof of
Theorem  \ref{min1}, the size of $L$ is zero. \qed

From Theorem \ref{claim3} and Theorem \ref{hardness_L}, we see that Algorithm \ref{algL2} approximates the problem 
when $L=O(\log n)$, which is also a $\log n$ hard problem, to the best possible extent unless $\NP \subseteq
\DTIME(n^{\log \log n})$.

From Theorem \ref{claim3} and since $L\subseteq N(p)$, it follows that if $d(p) = O(\log n)$ then the same 
algorithm gives an $O(\log n)$
approximate solution. As a corollary to Theorem \ref{claim3} we have the following result using Theorem \ref{max1}.

\begin{cor}
\texttt{MDD(min)} can be approximated within a factor of $O(\log n)$ provided $d(p)\geq n-O(\log n)$. 
\end{cor}


We now consider algorithms for \texttt{MDD(max)} on regular graphs. We arrive at the following Lemma:
\begin{lem}
Let $G=(V, E)$ be a $k$-regular graph with $|V|=n$ and $S$ be any solution to \texttt{MDD(max)} for $G$.  Let $(S, V \setminus S)$ be the set of edges across the sets $S$ and $V-S$ and $f = |N(p) \setminus S|$. Then $|S| \geq \frac{(k-f+1)n -1}{2k-f+1} \geq \frac{n-1}{k+1}$. \label{lemdel}
\end{lem}
\noindent
{\bf Proof:}
By using estimations on $|(S, V \setminus S)|$, we see that
\begin{equation}
k|S| \geq |(S, V \setminus S)| \geq (k-f) + (k-f+1)(n-|S| - 1). \label{eqdel}
\end{equation}
Note that the leftmost term represents the maximum number of edges that can arise out of $S$, and that the rightmost term is a lower bound on the number of edges arising out of $V\setminus S$. From \eqref{eqdel}, the proof of the Lemma follows. \qed

From Lemma \ref{lemdel}, we have
\begin{thm}
\texttt{MDD(max)} can be approximated within a factor of $(k+1)$ on $k$-regular graphs.
\end{thm}

However, it is possible to improve this approximation bound for \texttt{MDD(max)} on cubic graphs. 
For this, we use the algorithms for \texttt{MinDom}  and \texttt{MinDissoVD} (Minimum Dissociation Vertex Deletion) 
given by Halldorsson \cite{haldersson} and Tu and Yang \cite{tuyang}, respectively.
Dissociation number of a given graph is the size of a maximum induced sub-graph of $G$ whose maximum degree
is 1. \texttt{MinDissoVD} is the vertex deletion problem corresponding to 
Dissociation number - the minimum number (or weight)
of vertices to be deleted such that the remaining graph has maximum degree 1.

\begin{prop} {\rm \cite{haldersson}}\label{cubicdom}
\texttt{MinDom} on unweighted cubic graphs can be approximated within a factor of 1.583.
\end{prop}

\begin{prop}{\rm \cite{tuyang}}\label{cubicdiss}
\texttt{MinDissoVD} on unweighted cubic graphs can be approximated within a factor of 1.57.
\end{prop}

\begin{thm}\label{mddmax-cubic}
\texttt{MDD(max)} for unweighted cubic graphs can be approximated within a factor of 1.583.
\end{thm}
\noindent
{\bf Proof:} 
Let $S$ be a minimal solution to \texttt{MDD(max)} for $G$. It is easy to observe that 
if $d_{G[V\setminus S]}(p)=0$ then $S=V\setminus \{p\}$. Also, it is easy to observe that
for any feasible solution $S$ to $G$, $d_{G[V\setminus S]}(p)\ne 1$. There are only two other choices left for 
$d_{G[V\setminus S]}(p)$ which are 2 and 3. We shall try to find a solution in each of the cases and choose the 
smallest of these three kinds of solutions.

First we compute a solution $S$ to \texttt{MDD(max)} for $G$ such that $d_{G[V\setminus S]}(p)=3$. 
In this case, it is important to observe that $1 \leq |N(x) \cap (V \setminus N[p])| \leq 2$, for all $x \in N(p)$. 

We now construct the graph $G'$ from $G$ as follows. First, take a 
copy $G'$ of $G$. Remove $N[p]$ from $G'$. For each $x \in N(p)$ with exactly two neighbours $a$ and $b$ in 
$V\setminus N[p]$, we introduce two new vertices $x^1$ and $x^2$ and four new edges $(x^1, a), (x^1, b)$,
$(x^2, a), (x^2, b)$ into $G'$. For each vertex $x \in N(p)$ with exactly one neighbor $a$ in $V \setminus N[p]$,
we introduce exactly one new vertex $x^1$ and a new edge $(x^1, a)$ to $G'$. We shall refer to this resulting graph as
$G'=(V', E')$ and denote $X$ as the set of vertices that are added to the vertex set $V \setminus N[p]$. Let 
$X_v$ be the set of vertices which are introduced with respect to the vertex $v \in N(p)$, so that $V'=(V\setminus N[p])\cup X=(V\setminus N[p])\cup (\cup_{v\in N(p)}X_v)$. By construction, $1 \leq |X_v|\leq 2$ $\forall$ $v \in N(p)$.


Let $D'$ be a dominating set in $G'$. If $D' \cap X = \emptyset$, then it can be observed that $D'$ is a solution to 
\texttt{MDD(max)} for $G$ with $d_{G[V\setminus D']}(p)=3$. If $D' \cap X \neq \emptyset$, then we construct a 
set $D$ with $D \cap X = \emptyset$ and $|D| \leq |D'|$ as follows.
For any $v \in N(p)$ if $D' \cap X_v \ne \emptyset$, then replace $X_v \cap D'$ by $|X_v \cap D'|$ 
vertices from $N_{G'}(X_v)$, choosing vertices which were not already in $D'$. We shall denote the resulting 
new vertex set as $D$.   
Using the fact that one vertex from $N_{G'}(X_v)$ is enough to dominate the  vertices of $X_v$, 
we can conclude that this new vertex set $D$ is a dominating set for $G'$. We claim that $D$ is also a 
solution to \texttt{MDD(max)} for $G$. Now, since $D$ is a dominating set for $G'$, then every vertex in 
$V(G')\setminus D$ has atleast one neighbour in $D$. This implies that every vertex in $(V\setminus \{p\}) 
\setminus D$ has atleast one neighbour in $D$. This means that  $d_{G[V\setminus D]}(v)\leq 2$ for 
every $v\in V\setminus (D\cup \{p\})$, while  $d_{G[V\setminus D]}(p)=3$. Hence $D$ is a solution 
to \texttt{MDD(max)} for $G$.
Conversely, if $S$ is a solution to \texttt{MDD(max)} for $G$ with $d_{G[V\setminus D]}(p)=3$, 
 then $S$ is a dominating set for $G'$.

Suppose there exists a solution $S$ to \texttt{MDD(max)} for $G$ with $d_{G[V\setminus S]}(p)=2$.
Then the two neighbours of $p$ in $G[V\setminus S]$ (say $y$ and $z$) are not adjacent. It is also necessary that
$N(\{y, z\}) \setminus \{p\} \subseteq S$. Let $x \in N(p)\cap S$ and
let $X^*=N(p) \setminus \{x\}=\{y,z\}$. Now consider the graph $G^*=G[V^*]$ with  $V^*=V\setminus (N[X^*]\cup \{x\})$. 
If $T$ is a solution to \texttt{MinDissoVD} for $G^*$, then (it is easy to prove that) $T\cup \{x\} \cup N(X^*)$ is a
solution to \texttt{MDD(max)} for $G$. Conversely, if $S$  is a solution to \texttt{MDD(max)} for $G$ with
$d_{G[V\setminus S]}(p)=2$ and $x\in S\cap N(p)$,  then $S\setminus (\{x\}\cup N(X^*))$ is a solution to
\texttt{MinDissoVD} for $G^*$.

Using this idea, we give an algorithm for \texttt{MDD(max)} on cubic graphs as in Algorithm 
\ref{alg-mddmax-cubic}. Let $S_1 = V\setminus \{p\}$, $S_2= D_{opt}$ and $S_3 = T_{opt}$, where 
$D_{opt}$ and $T_{opt}$ are optimal solutions to \texttt{MinDom} for $G'$ and \texttt{MinDissoVD}  for $G^*$,
respectively. Then the set $S_{opt}$ defined as a smallest of $S_1, S_2$ and $S_3$
gives an optimal solution to \texttt{MDD(max)} for $G$. Conversely, if $S_{opt}$ is an optimal
solution to \texttt{MDD(max)} for $G$, then either $S_{opt}=S_1$, or $S_{opt}$ is an optimal solution to 
to \texttt{MinDom} for $G'$ or an optimal solution to \texttt{MinDissoVD} for $G^*$. 

\noindent
\begin{minipage}{1.\linewidth}
\begin{algorithm}[H]
\KwIn{A 3-regular graph $G=(V, E)$ and $p\in V$\;}
\KwOut{A solution $S$ to \texttt{MDD(max)}\;}
$S = V\setminus \{p\}$\;
\If{there is no $x\in N(p)$ such that $N[x]=N[p]$}{
Compute a dominating set $D$ for the graph $G'$ as in the proof of Theorem \ref{mddmax-cubic}\;
{\bf if~}{$|D| < |S|$} {\bf then} {$S=D$\;}
}
\For{each $x\in N(p)$}{
Let $N(p)-\{x\}=\{y, z\}$\;
\If {$(y, z)\notin E)$}{
Compute a solution $T$ to \texttt{MinDissoVD} for the input graph $G[V\setminus (\{x\}\cup
N[\{y, z\}])]$\;
}
$S'=T\cup \{x\}\cup N(y, z)$\;
{\bf if} $|S'| < |S|$ {\bf then} $S=S'$\;
}
return($S$)\;
\caption{\textbf{Computation of  1.583 factor solution to \texttt{MDD(max)} on cubic graphs}}
\label{alg-mddmax-cubic}
\end{algorithm}
\end{minipage}


Now, let $S$ be the solution returned by Algorithm \ref{alg-mddmax-cubic}. If $S_{opt}=D_{opt}$, then 
$ \dfrac{|S|}{|S_{opt}|} = \dfrac{|S|}{|D_{opt}|} \leq \dfrac{|D|}{|D_{opt}|}$,
where $D$ is the approximate solution to \texttt{MinDom} for $G'$. Then by Proposition \ref{cubicdom}, $S$ is an
approximate solution within a factor of 1.583.

Suppose $S_{opt}\cap N(p)=\{x\}$ and  $S_{opt}=T_{opt} \cup \{x\} \cup N(X^*)$. 
Let $T$ be an approximate solution to \texttt{MinDissoVD} for $G^*$. Let
$\alpha= |\{x\} \cup N(X^*)|$. Then we  have
$$ \dfrac{|S|}{|S_{opt}|} = \dfrac{|S|}{|T_{opt}| +\alpha} \leq \dfrac{|T| + \alpha}{|T_{opt}|+\alpha}
\leq \dfrac{|T|}{|T_{opt}|}.$$
Therefore, by Proposition \ref{cubicdiss}, $S$ is an approximate solution within a factor of 1.57. Hence,
the approximate solution returned by Algorithm \ref{alg-mddmax-cubic}
is within a factor of 1.583. \qed

\section*{Conclusion}
We have shown that both \texttt{MDD(min)} and \texttt{MDD(max)}, even when restricted to 
 bipartite graphs, cannot be approximated within a factor $O(\log n)$ unless $\NP \subseteq \DTIME(n^{\log \log
n})$. An approximation  within a factor of $O(\log n)$ is seen if $d(p)\leq O(\log n)$ or $d(p)\geq n-O(\log n)$
for \texttt{MDD(max)} and \texttt{MDD(min)}, respectively. Better approximation results for \texttt{MDD(min)} and
\texttt{MDD(max)} on bipartite graphs remain unknown and we conjecture that on general graphs, it
is hard to approximate both problems within a factor $O(2^{\log^{1-\epsilon}n})$, for any $\epsilon>0$.

\bibliographystyle{elsarticle-num}
\bibliography{research}

\begin{bibdiv}
\begin{biblist}

\bib{betzler}{incollection}{
      author={Betzler, Nadja},
      author={Bredereck, Robert},
      author={Niedermeier, Rolf},
      author={Uhlmann, Johannes},
       title={On making a distinguished vertex minimum degree by vertex
  deletion},
        date={2011},
   booktitle={Sofsem 2011: Theory and practice of computer science},
   publisher={Springer},
       pages={123\ndash 134},
}

\bib{betzler1}{article}{
      author={Betzler, Nadja},
      author={Uhlmann, Johannes},
       title={Parameterized complexity of candidate control in elections and
  related digraph problems},
        date={2009},
     journal={Theoretical Computer Science},
      volume={410},
      number={52},
       pages={5425\ndash 5442},
}

\bib{tnet2}{article}{
      author={Clauset, Aaron},
      author={Gleditsch, Kristian~Skrede},
       title={The developmental dynamics of terrorist organizations},
        date={2012},
     journal={PloS one},
      volume={7},
      number={11},
       pages={e48633},
}

\bib{terroristnet1}{article}{
      author={Clauset, Aaron},
      author={Moore, Cristopher},
      author={Newman, Mark~EJ},
       title={Hierarchical structure and the prediction of missing links in
  networks},
        date={2008},
     journal={Nature},
      volume={453},
      number={7191},
       pages={98\ndash 101},
}

\bib{protein2}{article}{
      author={Estrada, Ernesto},
       title={Virtual identification of essential proteins within the protein
  interaction network of yeast},
        date={2006},
     journal={Proteomics},
      volume={6},
      number={1},
       pages={35\ndash 40},
}

\bib{feige}{article}{
      author={Feige, Uriel},
       title={A threshold of ln n for approximating set cover},
        date={1998},
     journal={Journal of the ACM (JACM)},
      volume={45},
      number={4},
       pages={634\ndash 652},
}

\bib{haldersson}{incollection}{
      author={Halld{\'o}rsson, Magn{\'u}s~M},
       title={Approximating k-set cover and complementary graph coloring},
        date={1996},
   booktitle={Integer programming and combinatorial optimization},
   publisher={Springer},
       pages={118\ndash 131},
}

\bib{kannThesis}{thesis}{
      author={Kann, Viggo},
       title={On the approximability of np-complete optimization problems},
        type={Ph.D. Thesis},
        date={1992},
}

\bib{okunbarak}{article}{
      author={Okun, Michael},
      author={Barak, Amnon},
       title={A new approach for approximating node deletion problems},
        date={2003},
     journal={Information processing letters},
      volume={88},
      number={5},
       pages={231\ndash 236},
}

\bib{papaYa}{article}{
      author={Papadimitriou, Christos~H},
      author={Yannakakis, Mihalis},
       title={Optimization, approximation, and complexity classes},
        date={1991},
     journal={Journal of computer and system sciences},
      volume={43},
      number={3},
       pages={425\ndash 440},
}

\bib{protein3}{article}{
      author={Raman, Karthik},
       title={Construction and analysis of protein-protein interaction
  networks},
        date={2010},
     journal={Autom Exp},
      volume={2},
      number={1},
       pages={2},
}

\bib{tuyang}{article}{
      author={Tu, Jianhua},
      author={Yang, Fenmei},
       title={The vertex cover p3 problem in cubic graphs},
        date={2013},
     journal={Information Processing Letters},
}

\bib{protein1}{article}{
      author={Yu, Haiyuan},
      author={Kim, Philip~M},
      author={Sprecher, Emmett},
      author={Trifonov, Valery},
      author={Gerstein, Mark},
       title={The importance of bottlenecks in protein networks: correlation
  with gene essentiality and expression dynamics},
        date={2007},
     journal={PLoS computational biology},
      volume={3},
      number={4},
       pages={e59},
}

\end{biblist}
\end{bibdiv}

\end{document}